\documentclass{appolb}

\usepackage{epsfig}
\usepackage{amsfonts}
\usepackage{amsmath}
\usepackage{amssymb}
\usepackage{graphicx}
\usepackage{mathrsfs}
\usepackage{hhline,color}
\usepackage{pifont,ulem}
\usepackage[utf8]{inputenc}

\begin{document}
\title{Inverse magnetic catalysis in the  
Polyakov--Nambu--Jona-Lasinio and entangled Polyakov--Nambu--Jona-Lasinio models
\thanks{Presented at EEF70}%
}
\author{Constan\c{c}a Provid\^encia, M\'arcio Ferreira, Pedro Costa
\address{Centro de F\'{\i}sica Computacional, Department of Physics,
University of Coimbra, P-3004 - 516  Coimbra, Portugal}
}
\maketitle

\begin{abstract}
We investigate the QCD phase diagram at zero chemical potential and finite
temperature in the presence of an external magnetic field within the 
three flavor Polyakov--Nambu--Jona-Lasinio and entangled 
Polyakov--Nambu--Jona-Lasinio models looking for the inverse magnetic catalysis. 
Two scenarios for a scalar coupling parameter dependent on the
magnetic field intensity are considered. These dependencies of the coupling 
allow to reproduce qualitatively lattice QCD results for the quark condensates 
and for the Polyakov loop: due to the magnetic field the quark condensates 
are enhanced at low and high temperatures and suppressed for temperatures close 
to the transition temperatures while the Polyakov loop increases with the 
increasing of the magnetic field.
\end{abstract}

\PACS{24.10.Jv, 11.10.-z, 25.75.Nq}
  
\section{Introduction}

Presently, the investigation of magnetized quark matter is attracting the 
attention of the physics community due to its relevance for different regions 
of the QCD phase diagram \cite{{Marcio_outros}}: from the heavy ion collisions 
at very high energies, to the understanding of the early stages of the Universe 
and for studies involving compact objects like magnetars.
In the presence of an external magnetic field $B$, the competition between two 
different mechanisms determine the behavior of quark matter: on one hand, 
the increase of low energy contributions leads to an enhancement of the quark 
condensate; on the other hand, the suppression of 
the quark condensate due to the partial restoration of chiral symmetry.

At zero baryonic chemical potential, almost all low-energy effective models, 
including the Nambu--Jona-Lasinio (NJL)-type models, find an enhancement of the 
condensate due to the magnetic field, the so-called magnetic catalysis (MC), and 
no reduction of the pseudocritical chiral transition temperature with the 
magnetic field \cite{reviews}.
However, the suppression of the quark condensate, also known as inverse magnetic 
catalysis (IMC), was obtained in lattice QCD (LQCD) calculations with physical 
quark masses \cite{baliJHEP2012,bali2012PRD,endrodi2013}. Due to the IMC effect
the pseudocritical chiral transition temperature decreases and the Polyakov loop 
increases with increasing $B$. 
In Ref. \cite{endrodi2013} it is argued that the IMC may be a consequence of 
how the gluonic sector reacts to the presence of a magnetic field, and, it is 
shown that the magnetic field drives up the expectation value of the Polyakov 
field. The distribution of gluon fields changes as a consequence of the 
distortion of the quark loops in the magnetic field background. 
Therefore, the backreactions of the quarks on the gauge fields should be 
incorporated in effective models in order to describe the IMC.

It is also known that in the region of  low momenta, relevant for 
chiral symmetry breaking, there is a strong screening effect of the gluon 
interactions which suppresses the condensate \cite{endrodi2013,Miransky:2002rp}. 
In this region, the gluons acquire a mass $M_g$ of the order of 
$\sqrt{N_f \alpha_s|eB|}$, due to the coupling of the gluon field to a 
quark-antiquark interacting state. 
In the presence of a strong enough magnetic field, this mass $M_g$ for gluons
becomes larger.
This, along with the property that the strong coupling $\alpha_s$ decreases 
with increasing $B$ ($\alpha_s(eB)\sim[b\ln(|eB|/\Lambda_{QCD}^2)]^{-1}$ with 
$b=(11N_c-2N_f)/12\pi=27/12\pi$ \cite{Miransky:2002rp}), leads  to an effective 
weakening of the  interaction between the quarks in the presence of an external
magnetic field, and damps  the chiral condensate. 
This suggests that the effective interaction between the quarks should include
the reaction of the gluon distribution to the magnetic field background. 
Having this in mind, the present work shows two different approaches of taking 
into account the influence of the presence of an external magnetic field
in the gluonic sector. 

We perform our calculations in the framework of the Polyakov--Nambu--Jona-Lasinio 
(PNJL) model. The Lagrangian in the presence of an external magnetic field is 
given by
\begin{eqnarray}
{\cal L} = {\bar{q}} \left[i\gamma_\mu D^{\mu}-{\hat m}_f \right ] q ~+~ 
	G_s \sum_{a=0}^8 \left [({\bar q} \lambda_ a q)^2 + ({\bar q} i\gamma_5 \lambda_a q)^2 \right ]\nonumber\\
	-K\left\{{\rm det} \left [{\bar q}(1+\gamma_5) q \right] + 
	{\rm det}\left [{\bar q}(1-\gamma_5)q\right]\right\} + 
	\mathcal{U}\left(\Phi,\bar\Phi;T\right) - \frac{1}{4}F_{\mu \nu}F^{\mu \nu},
	\label{Pnjl}
\end{eqnarray}
where the quarks couple to a (spatially constant) temporal background gauge 
field, represented in terms of the Polyakov loop.
Besides the chiral point-like coupling $G_s$, that denotes the coupling of the 
scalar-type four-quark interaction in the NJL sector, in the PNJL  model the 
gluon dynamics is reduced to the chiral-point coupling between quarks together 
with a simple static background field representing the Polyakov loop. 
The Polyakov potential $\mathcal{U}\left(\Phi,\bar\Phi;T\right)$ is introduced 
and depends on the critical temperature $T_0$, that for pure gauge is 270 MeV.
In addition to the PNJL model, we also consider the effective vertex depending 
on the Polyakov loop \cite{EPNJL} (EPNJL model), 
$G_s(\Phi;\bar{\Phi})= G_s [1-\alpha_1\Phi\bar{\Phi}-\alpha_2(\Phi^3+\bar{\Phi}^3)]$,
that generates an entanglement interaction between the Polyakov loop and the 
chiral condensate.

\begin{figure}[t!]
\centering
	\includegraphics[width=0.49\linewidth,angle=0]{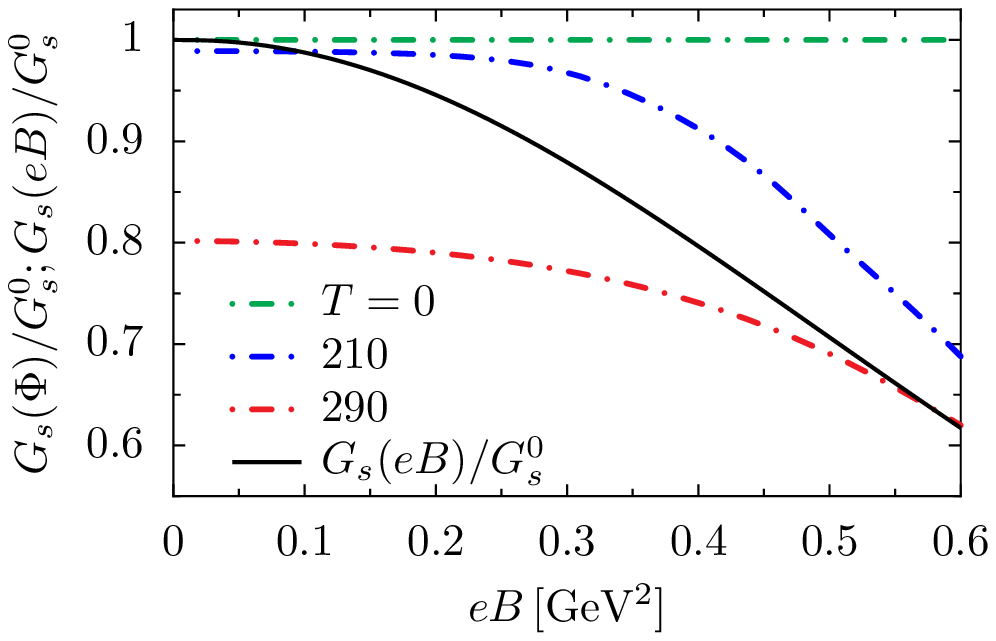}
  \includegraphics[width=0.49\linewidth,angle=0]{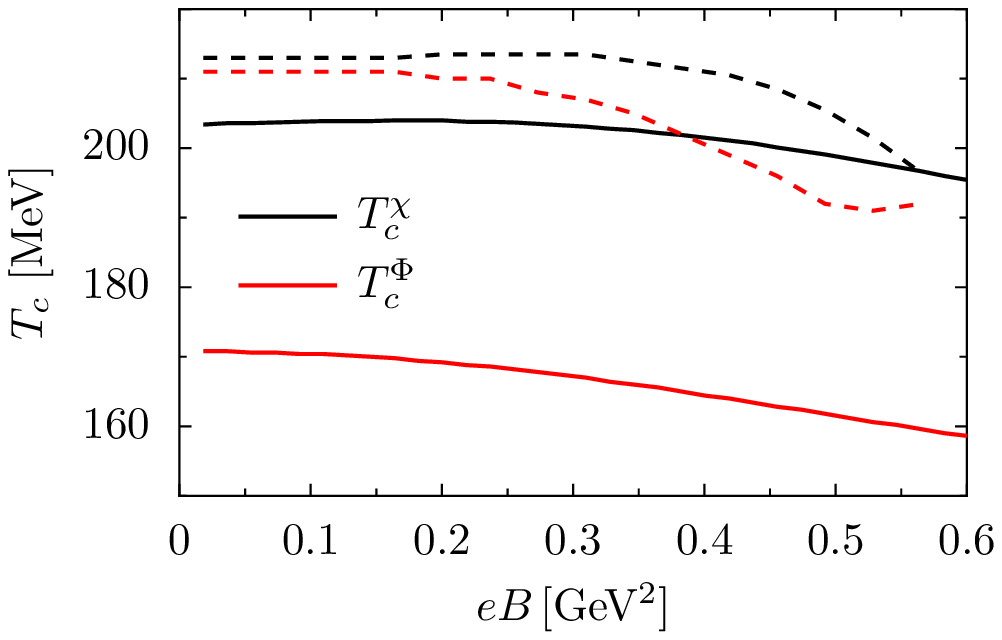}
  \caption{Comparison between both cases: left panel $-$ the scalar coupling 
			$G_s$ versus the magnetic field; 
			right panel $-$ the chiral and deconfinement temperatures as a function
			of $eB$, being the full (dashed) lines for {\it Case I} ({\it Case II}).
}
\label{fig:fit_Tc}
\end{figure}

In {\it Case I} we adopt a running coupling of the chiral invariant quartic 
quark interaction in the PNJL model with the magnetic field 
\cite{Farias:2014eca,MarcioIMC}.  
The damping of the strength of the effective interaction is build 
phenomenologically: since there is no available LQCD data for $\alpha_s(eB)$, 
we fit $G_s(eB)$ in order to reproduce the chiral pseudocritical temperature 
$T_c^\chi(eB)$ obtained in LQCD calculations \cite{baliJHEP2012}. 
The $G_s(eB)$ coupling, that reproduces $T_c^\chi(eB)$, is calculated in the NJL 
model and is shown in the left panel of Fig. \ref{fig:fit_Tc} (solid black line).
Now, using this $G_s(eB)$ coupling in the PNJL model, both the 
deconfinement transition and  chiral transition pseudocritical temperatures are 
decreasing functions with $eB$, up to $eB\sim 1$ GeV$^2$. 
Due to the existing coupling between the Polyakov loop field and quarks, the 
$G_s(eB)$ does not only affect the chiral transition but also the deconfinement 
transition. 

In {\it Case II} we introduce an $eB-$dependence on the pure-gauge critical 
temperature $T_0$, reproducing the LQCD data for the deconfinement transition 
\cite{baliJHEP2012}, in order to mimic the reaction of the gluon sector to the 
presence of an external magnetic field.

We will use the EPNJL model because within the PNJL it is not possible to 
implement the above scheme, since the chiral transition temperatures 
increase strongly with the external magnetic field. In order to bring these 
temperatures down, it would be necessary to use very small values of $T_0$, 
for which the deconfinement phase transition becomes of first order. 

Nevertheless, within EPNJL the chiral condensates and the Polyakov loop are 
entangled. Thus, the chiral transition temperatures are pulled down to 
temperatures close to the deconfinement transition temperature. This model, 
however, at moderate magnetic fields, still predicts a first order transition
for both transitions, when a small $T_0$ is needed. 
As a consequence, a too small value of $T_0$ leads to a 
first order phase transition within both PNJL and EPNJL models, and, therefore, 
the range of $T_0$ values we are interested in is limited to the values that 
maintain the crossover transition. 
A larger range of validity is obtained if the quark backreactions are not taken 
into account at $eB = 0$, i.e. when $T_0 = 270$ MeV as obtained in pure gauge. 
This gives $T^\Phi_c = 214$ MeV, 40 MeV higher than the prediction of lattice 
QCD data in \cite{baliJHEP2012}. 
This parametrization reproduces the referred lattice QCD data for $T_c^{\Phi}(eB)$, 
shifted by 40 MeV, for magnetic fields up to $0.61\,\mbox{GeV}^2$.
Above  0.61 GeV$^2$, a  first order phase transition occurs. 
We will use the last scenario in {\it Case II} to illustrate our results because 
larger magnetic fields are achieved.

Moreover, in the EPNJL the coupling $G_s$ depends on the Polyakov loop, thus, in
the crossover region, where the Polyakov loop increases with temperature, the 
coupling $G_s$ becomes weaker. This is shown in Fig. \ref{fig:fit_Tc} (left panel), 
where the coupling $G_s[\Phi(T)]$ is plotted for several temperatures 
(dashed curves) \cite{Ferreira:2013tba}. 
Within the PNJL model with constant coupling $G_s$, no IMC effect was obtained 
even with $T_0(eB)$, because $T_0(eB)$ does not affect the coupling $G_s$.

In Fig. \ref{fig:fit_Tc} (right panel), the results for the
pseudocritical temperatures for both cases are compared. 
In {\it Case II}, the pseudocritical temperatures have a much flatter behavior 
at small values of $eB$ than in {\it Case I}, reflecting the softer decrease of 
the coupling $G_s$ at small magnetic field values as shown in 
Fig. \ref{fig:fit_Tc} (left panel). Also, within the EPNJL with $T_0(eB)$, the 
difference between the pseudocritical temperatures $T_c^\chi$ and $T_c^\Phi$ 
is much smaller, due to the strong coupling between the Polyakov loop and the quark 
condensates. For $eB=0$ these temperatures are almost coincident, but a finite 
strong magnetic field destroys this coincidence. 
The PNJL model with $G_s(eB)$  does not have this feature and  different 
temperatures for  $T_c^\chi$ and $T_c^\Phi$ are predicted.

Next, we discuss the effect of the magnetic field on the quark condensates
and on the Polyakov loop, for both cases.

According to \cite{bali2012PRD}, we define the change of the light quark 
condensate due to the magnetic field as
$
\Delta\Sigma_{f}(B,T)=\Sigma_{f}(B,T)-\Sigma_{f}(0,T),
$
with
$
\Sigma_{f}(B,T)=\frac{2M_f}{m_\pi^2f_\pi^2}
\left[\left\langle \bar{q}_{f}q_{f}\right\rangle(B,T) \right.
- \left. \left\langle \bar{q}_{f}q_{f}\right\rangle(0,0) \right]+1,
$
where the factor $m_\pi^2f_\pi^2$ in the denominator contains the pion mass in 
the vacuum ($m_\pi=135$ MeV) and (the chiral limit of the) pion decay constant 
($f_\pi=87.9$ MeV) in NJL model. 

\begin{figure*}[t!]
    \includegraphics[width=0.49\linewidth,angle=0]{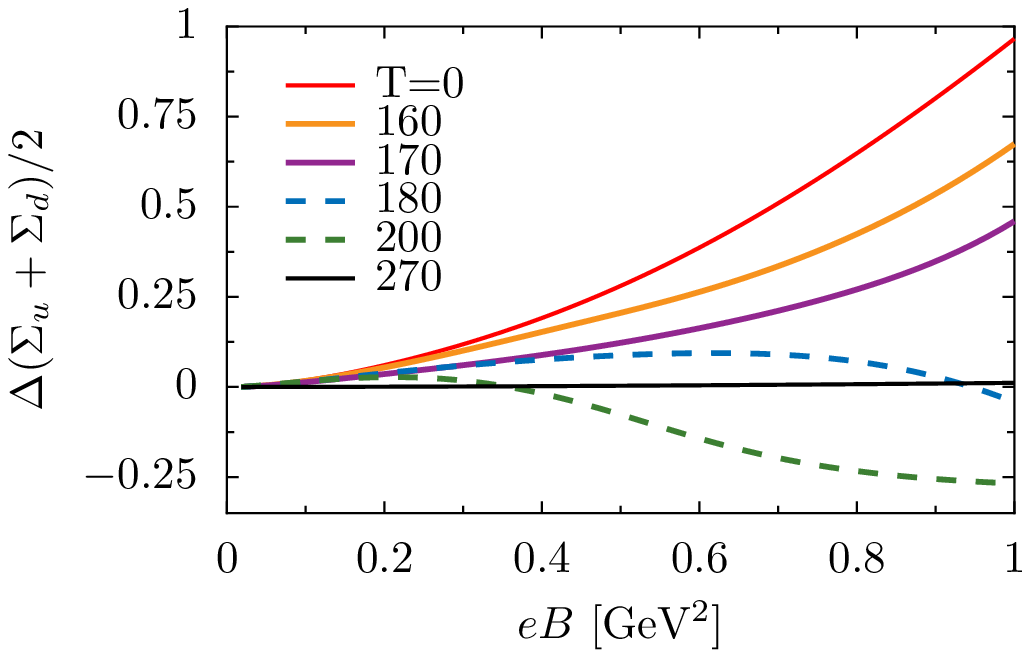}
    \includegraphics[width=0.49\linewidth,angle=0]{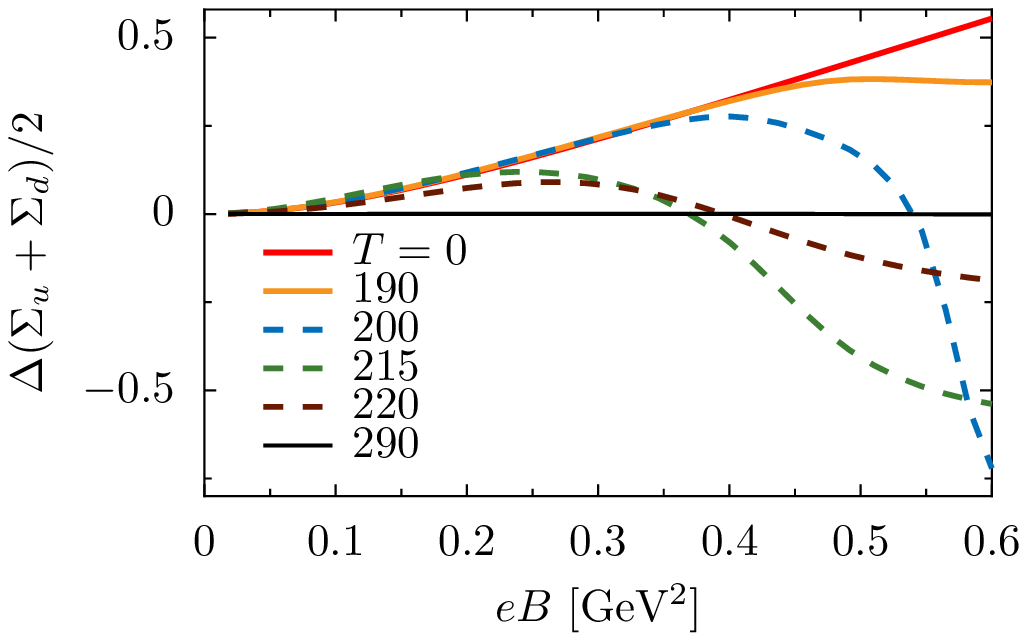}
    \caption{ 
    The light chiral condensate, $\Delta(\Sigma_u+\Sigma_d)/2$, as a function of 
		$eB$ for several values of $T$ in MeV: left panel $-$ {\it Case I};
		 right panel $-$ {\it Case II}.
    }
\label{fig:condensate}
\end{figure*}

In  Fig. \ref{fig:condensate} the light chiral condensate 
$\Delta(\Sigma_u+\Sigma_d)/2$ is plotted as a function of the magnetic field, 
for $eB<1$ ($eB<0.61$) GeV$^2$ in {\it Case I (Case II)}, at
temperatures close to the respective $T_c^{\Phi}(eB=0)$. 
The main conclusions are: 
{\bf i)} for both cases the qualitative behavior shown in Fig. 2 of Ref. 
\cite{baliJHEP2012} and in Fig. 6 of Ref. \cite{endrodi2013} is reproduced, 
that is, the non-monotonic behavior of the condensates as a function of the 
magnetic field, having the $T=0$ curves the highest $\Delta(\Sigma_u+\Sigma_d)/2$;
{\bf ii)} for temperatures close $T_c^{\chi}(eB=0)$ 
the strong interplay between the partial restoration of chiral 
symmetry and the condensate enhancement due to the magnetic field gives
rise to curves that increase, for small values of $eB$, and as soon as the partial 
restoration of chiral symmetry becomes dominant the curve starts to decrease. 

\begin{figure*}[t!]
    \includegraphics[width=0.49\linewidth,angle=0]{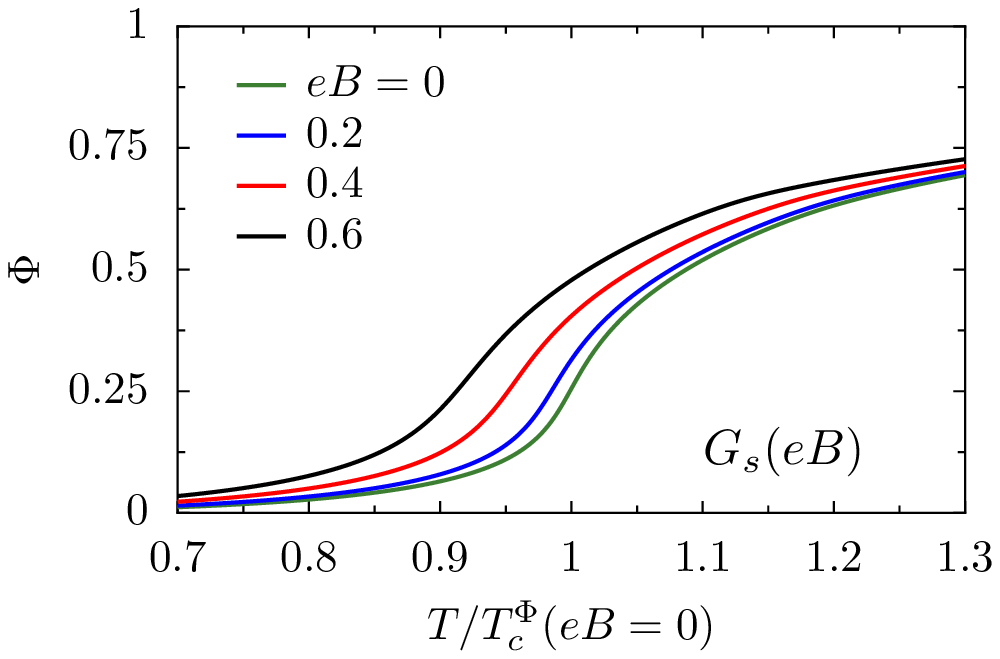}
    \includegraphics[width=0.49\linewidth,angle=0]{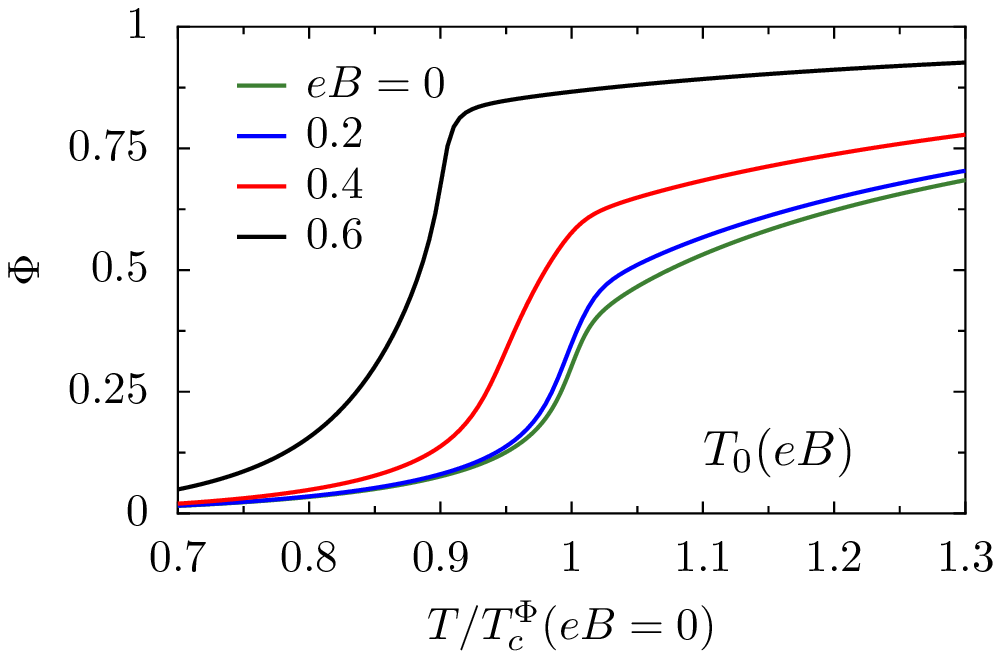}
    \caption{ 
		The Polyakov loop as a function of $T$ for different values of $eB$ (in GeV$^2$)
		renormalized by the deconfinement pseudocritical temperature at $eB = 0$: 
		 $T_c^{\Phi}= 171$ MeV for {\it Case I} (left panel) and $T_c^{\Phi}= 214$ MeV  
		for {\it Case II}, (right panel).
		}
\label{fig:polyakov}
\end{figure*}

Finally, the effect of the magnetic field on the Polyakov loop is seen in 
Fig. \ref{fig:polyakov}, where the Polyakov loop value, $\Phi$, is plotted as 
a function of the temperature, for several magnetic field strengths.
As can be seen, for both cases, the Polyakov loop increases sharply with 
the magnetic field around the transition temperature, and the transition
temperature decreases with the magnetic field with $B$,
in close agreement with the LQCD results \cite{endrodi2013}.
Indeed, the suppression of the condensates achieved by the magnetic field 
dependence of the coupling parameter results in an increase of the Polyakov loop, 
with this effect being stronger precisely for temperatures close to 
the transition temperature.

\vspace{0.25cm}
{\bf Acknowledgments}:
This work was partly supported by Project PEst-OE/FIS/UI0405/2014 
developed under the initiative QREN financed by the UE/FEDER through the
program COMPETE $-$ ``Programa Operacional Factores de
Competitividade'', and by Grant No. SFRH/BD/\-51717/\-2011.

\end{document}